\documentclass[doublecol]{epl2}	
\title{Emergence of a non-scaling degree distribution in\\ bipartite networks: a
numerical and analytical study}

\shorttitle{Non-scaling degree distribution in bipartite networks}

\author{
  Fernando Peruani \inst{1,2} \and
  Monojit Choudhury \inst{3} \and 
  Animesh Mukherjee \inst{3} \and 
  Niloy Ganguly \inst{3} }

\shortauthor{Peruani, Choudhury, Mukherjee and Ganguly}

\institute{
\inst{1} Max Planck Institute for Physics of Complex Systems,
  N\"othnitzer Strasse 38, 01187 Dresden, Germany \\
\inst{2} ZIH, TU Dresden, Zellescher Weg 12, 01069 Dresden,
  Germany, \\
\inst{3} Department of Computer Science and Engineering, Indian Institute of Technology Kharagur, India.
 }


\abstract{ We study the growth of bipartite networks in which the number of nodes in one of the partitions is kept fixed  while the other partition is allowed to grow. We
study random and preferential attachment as well as combination of both. We derive the exact
analytical expression for the degree-distribution of all these different types of attachments  while assuming that edges are  incorporated sequentially, i.e.,  a single edge is
added to the growing network in a time step. We also provide an approximate expression for the case when more than one edge are added in a time step. 
We show that depending on the relative weight between random 
and preferential  attachment, the degree-distribution of this type
of network falls into one of four possible regimes which range from
a binomial distribution for pure random attachment to an u-shaped
distribution for dominant preferential attachment.
}

\begin{document}

\maketitle

%%%%%%%%%%%%%%%%%%%%%%%%%%%%%%%%%%%%%%%%%%%%%%%%%%%%%%%%%%%%%%%%
%%%%%%%%%%%%%%%%%%%%%%Section 1 %%%%%%%%%%%%%%%%%%%%%%%%%%%%%%%%%%
%%%%%%%%%%%%%%%%%%%%%%%%%%%%%%%%%%%%%%%%%%%%%%%%%%%%%%%%%%%%%%%%%%

% \section{Introduction}

\begin{figure}
\centering\resizebox{\columnwidth}{!}{\rotatebox{0}{\includegraphics{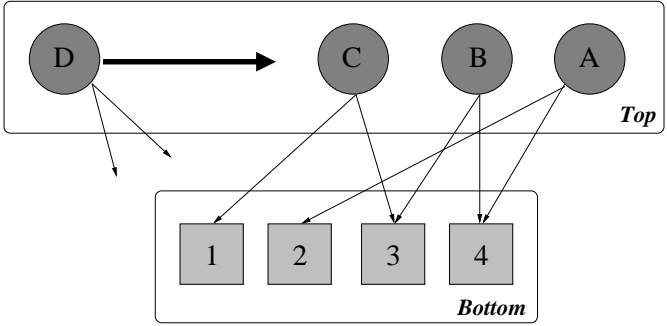}}}
\caption{Scheme of a growing bipartite network. In the example, the number of fixed $bottom$ nodes is $N=4$. 
Each $top$ nodes arrives with $\mu=2$  edges. The $top$ node $D$ represents a new incoming node.} \label{fig:scheme_bnw}
\end{figure}

A bipartite network is a graph which connects two distinct sets %%%KGP%%%%%%
(or partitions) of nodes, which we will refer to as the $top$ and the $bottom$ set. %%%KGP%%%%%%
%The edges
%of the network connect $top$ and $bottom$ nodes, but never two nodes
%of the same subset.
An edge in the network runs between a pair of a $top$ and a $bottom$ node but never between a pair of $top$ or a pair %%@
of $bottom$ nodes (see Fig. \ref{fig:scheme_bnw}).
Typical examples of this type of networks include collaboration
networks 
such as the movie-actor~\cite{ramasco,watts_98,collaboration3,Alava:06,movie1,movie2}, %%@
article-author~\cite{collaboration4,collaboration5,collaboration6,collab7,lambiotte_05}, and board-director~\cite{caldarelli_05, strogatz_01} network. 
In the movie-actor network, for instance, the movies and actors are
the elements of the $top$ and the $bottom$ set respectively, and an edge
between an actor $a$ and a movie $m$ indicates that $a$ has acted in
$m$. The actors $a$ and $a'$ are $collaborators$ if both have participated in the same movie, i.e., if both are connected to the same node $m'$.
The concept of $collaboration$ can be extended to include so diverse phenomena represented by bipartite networks as the city-people network~\cite{eubank_04}, in which an edge
between a person and a city indicates that the person has visited that particular city, the  
word-sentence~\cite{cancho,Latapy}, bank-company~\cite{souma_03} or donor-acceptor network, which accounts for injection and merging of magnetic field lines~\cite{sneppen_04}.

Several models have been proposed to synthesize the structure of
bipartite networks when both partitions grow unboundedly over time ~\cite{ramasco,watts_98,collaboration3,Alava:06,Latapy}. 
It has been found that for such growth models when each incoming $top$ node connects
through preferential attachment to $bottom$ nodes the emergent
degree distribution of $bottom$ nodes follows a power-law~\cite{ramasco}.
Another important property of bipartite networks is that the clustering coefficient cannot be measured in the standard way~\cite{watts_98}, and has to be measured as a cycle of four
connections ~\cite{lind_06}.

On the other hand, bipartite networks, where one of the partitions remains fixed over time (i.e., the number of 
$bottom$ nodes are constant), have received comparatively much less attention. 
Recently it was shown through numerical simulations that restrictions in the growth rate of the partitions can lead to non-scaling degree distribution highly sensitive to the parameters of
the growth model~\cite{dahui_05}. However, there is still no systematic and analytical study of this kind of networks.
Realizations of this type of bipartite networks include 
numerous relevant 
systems such as 
the interaction between the codons and genes as well as amino acids and proteins in biology and elements and compounds 
in chemistry.  
We can also include 
in this group those networks in which one partition can be
considered to be in a pseudo-steady state while the other one keeps on
growing at a much faster rate. For instance, it is reasonable to assume that for the city-people network~\cite{eubank_04}, the city growth rate is zero compared with the
population growth rate. Other examples of this type of networks could be the phoneme-language~\cite{Choudhury:06,Mukh:06}, in
linguistics, or train-station~\cite{sen_03,seaton} in logistics.
%

%Bipartite networks with fixed bottom come into play essentially 
%each time that we observe a system in which we can distinguish a set of
%building blocks and a set of entities that are made of them.
%To understand how these elements were combined to form these
%entities, and answer so relevant questions such as whether the
%element combination was performed randomly or through some sort of
%preferential scheme, a quantitative theory on the growth of this
%type of networks is needed

%%%%%%%%KGP%%%%%%%%%%%%%%%%
%We cannot claim that this is the only reason behind the formation of BNWs with fixed bottom. The issue being %%@
%debatable we might omit the above two paragraphs without any loss in the flow of the paper.
%%%% FP: I still believe that these paragraph that were deleted are the ones that could provide more relevance to this work. 

In this work, we study the growth of bipartite networks in which the
number of nodes in one of the partitions is kept fixed.
We explore random and 
preferential attachment as well as a combination of both.
We obtain an exact analytical expression for the degree distribution of the $bottom$ nodes assuming that the 
attachment is sequential, i.e., a single edge is added to the network in one time step. We also present an approximate 
solution for the case of parallel attachment, i.e., when more than one edge are incorporated into the network in a 
given time step. 
We show that, depending on the relative weight of random 
to preferential  attachment, the degree-distribution of this type
of network falls into one of four possible regimes which range from
a binomial distribution for pure random attachment to an u-shaped
distribution for dominant preferential attachment.
For combinations of random and preferential attachment the
degree-distribution asymptotically tends to beta-distribution with time.
%

%We find that these analytical expressions so obtained for the different conditions mentioned above 
%asymptotically tends to a beta-distribution with time.   

\section{The growth model}\label{sec:syn}

We consider the case in which the $top$ partition grows with time
while the number of nodes in the $bottom$ partition $N$ is kept
constant. We grow the network in the following way. At each time
step a new node is incorporated to the $top$ set.
Then, $\mu$ edges are connected from the new node to the nodes in
the $bottom$ set (see Fig. \ref{fig:scheme_bnw}).
The probability of attaching a new edge to the $bottom$ node $i$ is
$\widetilde{A}(k_{i}^{t})$, where $k_{i}^{t}$ refers to the degree of
the $bottom$ node $i$ at time $t$. We refer to
$\widetilde{A}(k_{i}^{t})$ as the attachment kernel and define it
as:
\begin{equation}\label{eq:kernel_attachment}
\widetilde{A}(k_{i}^{t}) = \frac{\gamma k_{i}^{t} + 1}{\sum_{j=1}^N
(\gamma k_{j}^{t} + 1)}
\end{equation}
where the sum in the denominator runs over all $bottom$ nodes, and
$\gamma$ is a model parameter which controls the relative weight of random to
preferential attachment. $\gamma$ can be
thought of as $\gamma=1/\alpha$, where  $\alpha$ is a positive constant
known in previous models %% for unbounded bottom partition %%KGP 
as {\em initial attractiveness}~\cite{mendes}.

There is a subtlety related to the attachment kernel and $\mu$ that
is worth to mention. The stochastic process can be performed in such
a way that the attachment of the $\mu$ incoming nodes is done
sequentially, i.e., one edge is attached per time step. This implies
that the denominator of $\widetilde{A}(k_{i}^{t})$ has to be updated
for each incoming node (and hence an edge), and that $t$ refers to the event of
incorporating a new edge to the $bottom$ set.
Alternatively, the attachment of the $\mu$ new edges can be done in
parallel. This implies that the new $\mu$ edges have all the same
probability of attaching to $bottom$ node $i$. In this case, $t$
refers to the event of incorporating a new node to the $top$ set.

There are two significant limits to consider: $\gamma = 0$ and
$\gamma \longrightarrow \infty$. For $\gamma = 0$, Eq.
(\ref{eq:kernel_attachment}) reduces to
$\widetilde{A}(k_{i}^{t})=1/N$, which implies that all $bottom$ nodes
have the same probability of being selected by an incoming edge.
This limit corresponds to pure random attachment.
For $\gamma \longrightarrow \infty$ Eq. (\ref{eq:kernel_attachment})
reduces to $\widetilde{A}(k_{i}^{t})=k_{i}^{t}/\sum_{j=1}^N (
k_{j}^{t})$, which means that higher degree $bottom$ nodes have higher
probability of being selected. This case corresponds to pure
preferential attachment.

Stochastic simulations have been performed with the initial
condition 
where all $bottom$ nodes at time $t=0$ have zero degree, i.e., initially no
edges are connected to the $bottom$ nodes.
%
%%Clearly using such an initial condition pure preferential
%%attachment cannot be realistically modeled (since certain initial selections are random), but the behavior of the
%%system as $\gamma \longrightarrow \infty$ can be extrapolated.

\section{Evolution equation for sequential attachment}\label{sec:derive}

Now we aim to derive an evolution equation for the degree
distribution of the $bottom$ nodes. We focus on sequential attachment.
Let $p_{k,t}$ be the probability of finding a randomly chosen $bottom$
node with degree $k$ at time $t$.
We recall that $t$ refers to the $t$-edge attachment event.
$p_{k,t}$ can be defined as $p_{k,t} = \left< n_{k,t} \right>/N$,
where $n_{k,t}$ refers to the number of nodes in the $bottom$ set with
degree $k$ at time $t$, and $\left< ... \right>$ denotes ensemble
average, i.e. average over realizations of the stochastic process
described above.
We express the evolution of $p_{k,t}$ in the following way:
\begin{equation}\label{eq:markovchain}
p_{k,t+1} = (1 - A(k,t)) p_{k,t} + A(k-1,t) p_{k-1,t}
\end{equation}

where $A(k,t)$ refers to the probability that the incoming edge
lands on a $bottom$ node of degree $k$.
$A(k,t)$ can be easily derived from Eq. (\ref{eq:kernel_attachment}) and takes the
form: 
\begin{equation}\label{eq:kernel_mu1}
A(k,t)=\frac{\gamma k + 1}{\gamma t + N}
\end{equation}
The reasoning behind Eq. (\ref{eq:markovchain}) is the following. 
The probability of finding a $bottom$ node with degree $k$ at time
$t+1$ decreases due to the number of nodes, which have a degree $k$ at time $t$
and receive an edge at time $t+1$ therefore acquiring degree
$k+1$, i.e., $A(k,t) p_{k,t}$. Similarly, this probability increases due to the number of
nodes that at time $t$ have degree $k-1$ and receives an edge at time $t+1$
to have a degree $k$, i.e., $A(k-1,t) p_{k-1,t}$. Hence the net increase in the probability can be expressed as in Eq. 
(\ref{eq:markovchain}).

According to what was done in the stochastic simulations, we assume that at time $t=0$ all
$bottom$ nodes have zero degree, which implies the initial condition
$p_{k,t=0}=\delta_{k,0}$, where $\delta_{k,0}$ is the Kronecker
delta function.

\section{Exact solution for sequential attachment}\label{sec:derive}

We look for the exact analytical solution of sequential attachment.
For this purpose we express Eq. (\ref{eq:markovchain}) as:
\begin{equation}\label{eq:markovvectorial}
\mathbf{p}_{t+1} = \mathbf{M}_t\mathbf{p}_t = \big[\prod_{\tau=0}^t
\mathbf{M}_{\tau}\big] \mathbf{p}_0
\end{equation}
where $\mathbf{p}_t$ denotes the degree distribution at time $t$ and
is defined as $\mathbf{p}_t = [p_{0,t}\ p_{1,t}\ p_{2,t}\
\ldots]^T$, $\mathbf{p}_0$ is the initial condition expressed as
$\mathbf{p}_0 = [1\ 0\ 0\ \ldots ]^T$, and $\mathbf{M}_{\tau}$ is
the evolution matrix at time $\tau$ which is defined as:
\begin{equation}
\mathbf{M_{\tau}} = \left( \begin{array}{c c c c c}
1 - A(0, \tau) & 0 & 0 & 0 & \ldots\\
A(0, \tau) & 1 -A(1, \tau) & 0 & 0 & \ldots\\
0 & A(1, \tau)& 1 - A(2, \tau) & 0 & \ldots\\
\vdots & \vdots & \vdots & \vdots & \ddots
\end{array}
\right)
\end{equation}

Since our initial condition is a vector with zeros in all position
except in the first one, all the relevant information, i.e., the
degree distribution of the $bottom$ nodes, is in the first column of
$\big[\prod_{\tau=0}^t \mathbf{M}_{\tau}\big]$. A close inspection
to the evolution of this column, explicitly using Eq. (\ref{eq:kernel_mu1}), reveals that the $k$-th element of it, which corresponds
to $p_{k,t}$, can be expressed as:
\begin{equation}\label{eq:solutionsequential}
p_{k,t} = \left( \begin{array}{c} t \\ k \end{array}\right)
\frac{\prod_{i=0}^{k-1}{(\gamma i+ 1)} \prod_{j=0}^{t-1-k}{(N - 1 +
\gamma j)}}{\prod_{m=0}^{t-1}{(\gamma m + N)} }
\end{equation}
for $k \leq t$, and $p_{k,t} = 0$ for $k > t$, and where we have
defined  $\prod_{i=0}^{-1}{(...)}=1$, and $\left( \begin{array}{c} t \\ k \end{array}\right)$ refers to the combinatorial number $t!/[\left(t-k\right)! k!]$.

Eq. (\ref{eq:solutionsequential}) is the exact solution of Eq.
(\ref{eq:markovchain}) using as initial condition
$p_{k,t=0}=\delta_{k,0}$, i.e., the analytical expression of the
degree distribution of the $bottom$ nodes when sequential attachment
is applied.

In the limit of $\gamma=0$ Eq. (\ref{eq:solutionsequential}) reduces
to:
\begin{equation}\label{eq:solutionsequential_ra}
p_{k,t} = \left( \begin{array}{c} t \\ k \end{array}\right)
\left(\frac{1}{N}\right)^{k} \left(1-\frac{1}{N}\right)^{t-k}
\end{equation}
for $k \leq t$, and $p_{k,t} = 0$ for $k > t$. In other words, Eq.
(\ref{eq:solutionsequential_ra}) is the solution of the sequential
problem when pure random attachment is applied.
%
%% KGP Probably this is not required.
%Comparing (\ref{eq:solutionsequential_ra}) with Eq.
%(\ref{eq:solutionsequential}) it becomes evident that
%(\ref{eq:solutionsequential}) is a binomial distribution with an
%explicit time dependence.
%
\begin{figure}
\centering\resizebox{\columnwidth}{!}{\rotatebox{0}{\includegraphics{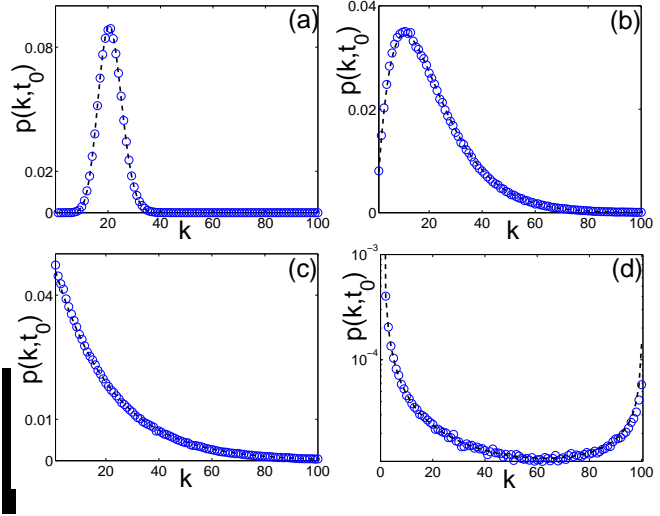}}}
\caption{The four possible degree distributions depending on $\gamma$. 
Symbols represent average over $5000$, in (a)-(c), and $50000$, in (d), stochastic simulations. The dashed curve is the theory given by Eq. (\ref{eq:solutionparallel}). 
From (a) to (c), $t_0=1000$, $N=1000$ and $\mu=20$. 
(a) at $\gamma=0$, $p(k,t)$ becomes a binomial distribution.
(b) $\gamma=0.5$, the
distribution exhibits a maximum which shifts with time for $0\leq\gamma < 1$.
(c) $\gamma=1$, $p(k,t)$ does not longer
exhibit a shifting maximum and the distribution is a monotonically
decreasing function of $k$ for $1\leq \gamma\leq (N/\mu) - 1$.
(d) $\gamma=2500$, $t_0=100$, $N=1000$ and $\mu=1$. $p(k,t)$ becomes an u-shaped curve for $\gamma > (N/\mu) - 1$.
} \label{fig:transition_snapshots}
\end{figure}
\section{Parallel attachment}\label{sec:derive}
We focus on parallel attachment, i.e., when more than one edge are
added per time step. We do not aim to derive an exact analytical
expression for the degree distribution of this problem but provide a
reasonable approximation.
We recall that for parallel attachment $t$ refers to the event of
incorporating a new $top$ node.
We assume that $\mu \ll N$ and expect Eq. (\ref{eq:markovchain}) to
be a good approximation of the process after replacing $A(k,t)$ with
$A_p(k,t)$. We define $A_p(k,t)$ as 
\begin{equation}\label{eq:kernel_mu_gt_1}
A_p(k,t)=\frac{\left( \gamma k + 1 \right)\mu}{\gamma \mu t + N} \, .
\end{equation}
The term $\mu$ in the denominator appears since in this case the total degree of the $bottom$ nodes at any point in time 
is $\mu t$ rather than $t$ as in Eq. (\ref{eq:kernel_mu1}). The numerator contains a $\mu$ since at each time step there are $\mu$ 
edges that are being incorporated into the network rather than a single edge.  

It is important to mention here that Eq. (\ref{eq:markovchain}) cannot exactly represent the stochastic
parallel attachment because it explicitly assumes that in one time
step a node of degree $k$ can only 
get converted to a node of degree $k+1$. Clearly,
the incorporation of $\mu$ edges in parallel allows the possibility
for a node of degree $k$ 
to get converted to a node of degree $k+ \mu$. The correct
expression for the evolution of $p_{k,t}$ reads:
\begin{equation}\label{eq:parallel}
p_{k,t+1} = (1 - \sum_{i=1}^{\mu}{\widehat{A}(k,i,t)}) p_{k,t} +
\sum_{i=1}^{\mu}{\widehat{A}(k-i,i,t)} p_{k-i,t}
\end{equation}
where $\widehat{A}(k,i,t)$ represents the probability at time $t$ of a node of
degree $k$ of receiving $i$ new edges in the next time step. We
expect Eq. (\ref{eq:markovchain}) to be a good approximation of Eq.
(\ref{eq:parallel}) when $\widehat{A}(k,1,t) \gg \widehat{A}(k,i,t)$ where $i>1$.

The solution of Eq. (\ref{eq:markovchain}) with the attachment kernel given by Eq. (\ref{eq:kernel_mu_gt_1})
 reads:
\begin{equation}\label{eq:solutionparallel}
p_{k,t} = \left( \begin{array}{c} t \\ k \end{array}\right)
\frac{\prod_{i=0}^{k-1}{\left(\gamma i+ 1\right)}
\prod_{j=0}^{t-1-k}{\left( \frac{N}{\mu} - 1 + \gamma
j\right)}}{\prod_{m=0}^{t-1}{\left(\gamma m + \frac{N}{\mu}\right)}
}
\end{equation}

We expect Eq. (\ref{eq:solutionparallel}) to approximate the degree
distribution of the stochastic process with parallel attachment for
$\mu \ll N$. This means that we cannot expect the approximation to hold for large values of $\gamma$ or $\mu/N$.

In the limit of random attachment, i.e., $\gamma=0$, Eq.
(\ref{eq:solutionparallel}) becomes $p_{k,t} = \left(
\begin{array}{c} t \\ k \end{array}\right)
\left(\frac{\mu}{N}\right)^{k} \left(1-\frac{\mu}{N}\right)^{t-k} $.

\begin{figure}
\centering\resizebox{\columnwidth}{!}{\rotatebox{0}{\includegraphics{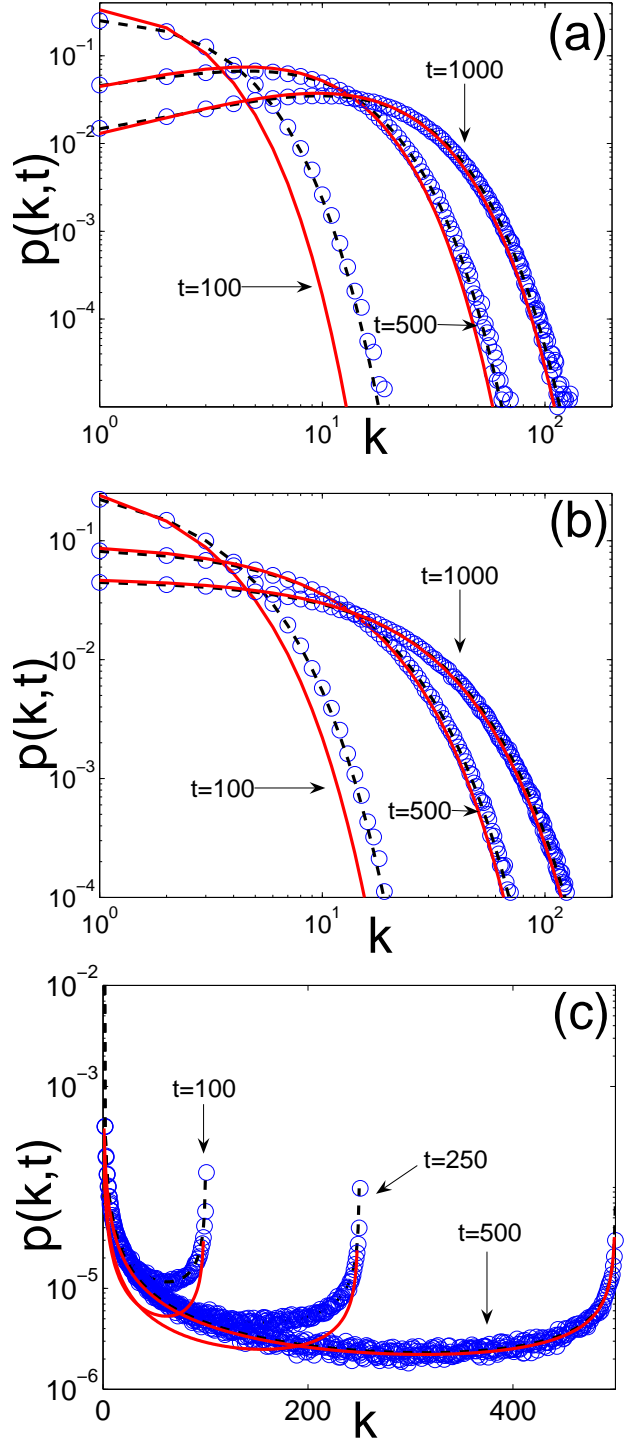}}}
\caption{Temporal evolution of $p(k,t)$ for various values of
$\gamma$. Symbols represent average over $5000$, in (a)-(b), and $50000$, in (c), stochastic
simulations. The black dashed curve is the theory, represented through Eq. (\ref{eq:solutionparallel}). 
The red solid curves correspond to the approximation given by the beta-distribution, Eq. (\ref{beta}). 
(a) $\gamma=0.5$, and (b) $\gamma=1$. $N=1000$ and $\mu=20$. Compare with Fig. \ref{fig:transition_snapshots}(b) and (c). 
(c) $\gamma=2500$, $N=1000$ and $\mu=1$, see Fig. \ref{fig:transition_snapshots}(d). }
\label{fig:temporalevolution}
\end{figure}

Figs. \ref{fig:transition_snapshots}(a)-(c) and \ref{fig:temporalevolution}(a)-(b) show a comparison between stochastic simulations and Eq.(\ref{eq:solutionparallel}) and prove that Eq. (\ref{eq:solutionparallel}) is a good approximation for 
$\mu \ll N$ and relative low values of $\gamma$. For large values of $\gamma$, as said above, the approximation fails. However, for $\mu=1$ Eq. (\ref{eq:solutionparallel})
reduces to Eq. (\ref{eq:solutionsequential}), which in this case is the exact solution, and then the theory works for all values of $\gamma$ (see Figs. \ref{fig:transition_snapshots}(d) and \ref{fig:temporalevolution}(c)).

\section{From random to preferential attachment}\label{sec:derive}

Fig. \ref{fig:transition_snapshots} shows that there is a clear
transition from random to preferential attachment. 
%
%Since Eq.
%(\ref{eq:solutionsequential}) and Eq. (\ref{eq:solutionparallel})
%exhibit the same qualitative behavior, we focus on the latter to
%perform the analysis of this transition.
%
At $\gamma=0$ (see Fig. \ref{fig:transition_snapshots}(a)) we observe that $p_{k,t}$ is centered around the
maximum (mode of the distribution) which shifts with time at a speed
of $\mu/N$ per time step, while the width of the distribution also spreads with
time. This behavior
corresponds to a situation in which all $bottom$ nodes receive roughly
the same amount of edges with time. The well defined maximum tells
us about the average number of edges each $bottom$ node has, while the 
variance of the distribution indicates the presence of fluctuations
around that mean value which increases with time.

For $0 < \gamma < 1$ the 
distribution is not longer symmetric (see Fig. \ref{fig:transition_snapshots}(b)). $Bottom$ nodes having small degree rarely receive an edge, and so  
$p_{k,t}$ decays slowly for small value of $k$. However the distribution still
exhibits a maximum, mode of the distribution, which shifts with time (see Fig. \ref{fig:temporalevolution}(a)).

For $1 \leq \gamma \leq (N/\mu)-1$ the distribution looses the (local) 
maximum and becomes monotonically decreasing (see Fig.
\ref{fig:transition_snapshots}(c)). We can always find a
$bottom$ node with small degree because small degree nodes hardly get
an edge. On the other hand, there are very few nodes with high
degree, and these ones receive almost all incoming edges. The temporal evolution of the distribution for this range of $\gamma$ is shown in Fig. \ref{fig:temporalevolution}(b).

For $\gamma > (N/\mu)-1$ the distribution described by Eq. (\ref{eq:solutionparallel})
exhibits an u-shape. As said above, we cannot expect Eq. (\ref{eq:solutionparallel}) to
approximate the stochastic process for such large values of $\gamma$. Stochastic simulations performed in this range of $\gamma$ for $\mu>1$ are very noisy and the u-shape
cannot be obtained by averaging over few simulations. However, for $\mu=1$ we can illustrate the u-shaped distribution in a clear way, see Fig.
\ref{fig:transition_snapshots}(d).  
As in the previous case, we still can find a
$bottom$ node with small degree because small degree nodes hardly get an edge (see maximum at $k=0$). But on the other hand, we can be sure that there is at least one node with
very large degree, because 
the node with the largest degree at time $t-1$ very likely is going to get an edge at time $t$, in an effect like ``winner takes all" (see peak at $k=t$). The node
with largest degree keeps on increasing its degree with time, and so $p_{k,t}$ has two peaks, one located at $k=0$ and the other one at $k=t$  (see Fig.
\ref{fig:temporalevolution}(c)).

\section{Beta Distribution}\label{sec:beta}
 
In the following we offer a  quantitative 
 analysis of the transition by showing that $p_{k,t}$ behaves asymptotically with
time as a beta-distribution for $\gamma > 0$.

For $t \gg \eta$, where $\eta = N/ (\gamma \mu)$, we can approximate
the products in Eq. (\ref{eq:solutionparallel}) by gamma-functions
and apply Stirling's approximation. After some algebra we obtain:
\begin{equation}\label{beta}
p_{k,t} \simeq  C^{-1}\left( k/t \right)^{\gamma^{-1}-1} \left(1-
k/t \right)^{\eta-\gamma^{-1}-1}
\end{equation}
where $C$ is a normalization constant defined by $C = \int_{0}^{t}{
\left( k'/t \right)^{\gamma^{-1}-1} \left(1- k'/t
\right)^{\eta-\gamma^{-1}-1}}dk'$.

Fig. \ref{fig:temporalevolution} shows a comparison between
stochastic simulations (circles), the theoretical solution given by Eq. (\ref{eq:solutionparallel}) (black dashed curve), and the approximation given by Eq. (\ref{beta}) (red
solid curve) for two different values of $\gamma$ at various times. In Fig.
\ref{fig:temporalevolution} it can be observed that $p_{k,t}$
approaches asymptotically to Eq. (\ref{beta}) (compare the black dashed and the red solid curves). Notice that Eq.
(\ref{beta}) does not have any fitting parameter and represents a
beta-distribution $f(x;\alpha, \beta)$ of the variable $x=k/t$ and
fixed parameters $\alpha=\gamma^{-1}$ and $\beta=\eta -
\gamma^{-1}$.

For $0<\gamma<1$,  $\alpha>1$ and $\beta>1$ the mode of the
distribution is given by $(\alpha - 1)/(\alpha + \beta
-2)=((\gamma^{-1})-1)/(\eta - 2)$. This can be easily verified by
taking the first derivative of Eq. (\ref{beta}) equal to zero. From
this we learn that in this range of $\gamma$ the maximum of the
distribution $k_{max}$ is located at
$k_{max}=t((\gamma^{-1})-1)/(\eta - 2)$ (see Fig.
\ref{fig:temporalevolution}(a)). In the limit of $\gamma
\longrightarrow 0$ we retrieve the behavior of $k_{max}$ observed
for pure random attachment, i.e., $k_{max}=t(\mu/N)$. 

At $\gamma=1$, $\alpha=1$ and $\beta>0$, the moving peak is no longer found,
i.e., the mode of the distribution is located for all times at
$k_{max}=0$ (see Fig.
\ref{fig:temporalevolution}(b)). This condition also holds for $1<\gamma\leq(N/\mu)-1$.

For $\gamma>(N/\mu)-1$, there is another regime for the degree distribution. For $\alpha<1$
and $\beta<1$, $p_{k,t}$ becomes u-shaped with a peak fixed at $k=0$ and the other one shifting with $t$. For $\mu=1$
the additional peak is located at $k=t$ (see Fig.
\ref{fig:temporalevolution}(c)).

\section{Concluding remarks}\label{sec:conc}

We have studied the growth of bipartite networks in which the number
of nodes in the $bottom$ set is kept fixed.
We consider random and preferential node attachment as well as the
combination of both.
We have derived the degree distribution evolution equation for
sequential and parallel attachment of nodes.
For sequential attachment we have provided the exact analytical
solution of the problem.
For parallel attachment we have obtained an approximate expression
for the degree distribution. Through simulations we have provided
numerical evidence which shows that the approximation for parallel attachment is reasonable
when $\mu \ll N$ and $\gamma$ is small.

Finally, we have shown that for both, sequential and parallel
attachment, the degree-distribution falls into one of four possible
regimes: a) $\gamma=0$, a binomial distribution whose mode shifts
with time, b) $0<\gamma<1$, a skewed distribution which
exhibits a mode that shifts with time, c) $1 \leq \gamma \leq
(N/\mu) - 1$, a monotonically decreasing distribution with the mode
frozen at $k=0$, and d) $\gamma>(N/\mu) - 1$, an u-shaped
distribution with peaks at $k=0$ and $k=t$. 
%
%%%%KGP NG --- says that the language example is more suitable since we have a better understanding of the same%%%
%Our results might help to understand the growth of bipartite
%networks as the codon-gene network by looking at the degree
%distribution at different times. In the particular example of
%codon-gene this can be done by looking at the degree distribution %% of
%different organisms far away separated in the evolution line. %%KGP
%for different organisms that would then be separated far away from each other on the time evolution line. The %%@
%analysis here presented might shed some light about the possible randomness behind the evolution of such processes.

Our results might be useful to explain the dynamical growth of various systems like the speech sound  
inventories of the world's languages, which can be rendered a bipartite structure as explained through the 
phoneme-language network in~\cite{Choudhury:06,Mukh:06}. A detailed study of the parameter $\gamma$ leading to the 
degree distribution of the network can then shed some light on the amount of randomness/preference that has gone into 
shaping the evolution of such a system.

%%%%%%%%%%%%%%%%%%%%%%%%%%%%%%%%%%%%%%%%%%%%%%%%%%%%%%%%%%%%%%%%%%%%%%
%%%%%%%%%%%%%%%%%%%%%%%%%%%%%%%%%%%%%%%%%%%%%%%%%%%%%%%%%%%%%%%%%%%%%%%

\acknowledgments

This work was supported by the Indo-German (DST-BMBT) project grant. The authors would like to 
extend their gratitude to Dr. Andreas Deutsch, Dr. Lutz Brusch and Dr. Luis G. Morelli for
their valuable comments and suggestions. MC and AM would like to thank Media Lab Asia and
Microsoft Research India respectively for financial assistance. They would also like to
extend their gratitude to Prof. Anupam Basu for providing them with the laboratory
infrastructure. FP would also like to acknowledge the hospitality of IIT-Kharagpur and financial support through Grant No. 
11111.

\end{document}